連続型スノードリフトゲームにおける協力とインセンティブ行動の進化
# Evolution of cooperation and costly incentives in the continuous snowdrift game

佐々木　達矢

ウィーン大学・数学

Tatsuya Sasaki

Faculty of Mathematics, University of Vienna

tatsuya.sasaki@univie.ac.at

**Abstract:** Game theory research on the snowdrift game has showed that gradual evolution of the continuously varying level of cooperation in joint enterprises can demonstrate evolutionary merging as well as evolutionary branching. However, little is known about the consequences of changes in diversity at the cooperation level. In the present study I consider effects of costly rewards on the continuous snowdrift game. I show that not evolutionary merging but evolutionary branching can promote the emergence of pool reward, which can then enhance social welfare.

## 1 Introduction

In contrast to the Prisoner's Dilemma, the snowdrift game can lead to a coexistence of cooperators and free riders [1]. The divergence scenario for the cooperation level in the continuous snowdrift game (SDG) has been termed the "tragedy of the commune" [2]: gradual evolution can favor such a state in which a sense of fairness may be minimized rather than a state in which all adopt the same cooperation level. Viewed in this light, of much interest would be to investigate the question: whether or not that evolutionary branching can make positive effects on the population's welfare? To date, the elementary classification of adaptive dynamics has been well provided for models with simple payoff functions [2,3], in which mostly only the negative conclusion of decrease in a sense of fairness has been considered. To my knowledge, the question has never seriously been tackled, despite the fact that coexistence of clearly different levels of cooperation is common in nature and human societies.

Here we approach the issue with considering pool reward as a facilitator of cooperation. It is well known that selective incentives such as reward and punishment are common tools to curb human behaviors [4,5]. In a standard pool-reward model based on discrete public good games, each player is offered the opportunities to contribute, first, to a fund pool that rewards cooperative behaviors, and second, to a pool that provides public goods; finally, the resulting goods are equally shared among all players, but the resulting rewards, among only those who contributed to the public goods [6,7]. Therefore, pool reward is a club good, not public, and its amount shared may depend on the level of altruism.

Voluntarily rewarding incurs some cost on each contributor. This thus can give rise to a dilemmatic situation among those who contribute to both ("pro-socialists") and those who contribute to the public good but not to the reward fund ("second-order free riders"). Recent results have shown, nevertheless, that voluntary reward for the pool system can evolve and improve cooperation and payoffs in discrete public good games, despite the presence of second-order free riders. Therefore, this implies that similarly, in the continuous public good game voluntary reward may thrive and then facilitate the social welfare, if a population was to be diversified as in the discrete case.

In the present study we focus on the SDG, a simple two-player public good game [2]. In what follows we investigate the gradual co-evolution of the investment level in the SDG and the donation level in the reward fund. We examine the model in particular for a range of parameter settings in which the continuous SDG leads initially monomorphic populations to undergo evolutionary branching on the investment level.

## 2 Model and methods

We consider an infinitely large, well-mixed population. Each individual $i$ in the population has two continuously varying traits, $(c_i, e_i)$ with $c_i, e_i \geq 0$, in which $c_i$ represents the degree of investment to the SDG, and $e_i$, the degree of donation in the reward fund. From time to time, a couple of players are randomly sampled from the population. Each player $i$ (= 1, 2) first invests $e_i$ to the reward fund at personal cost $S(e_i)$ and then invests $c_i$ to the public good at personal cost $C(c_i)$. Finally, each receives (i) the benefits from the SDG and (ii) the rewards from the reward fund. (i) The individual benefit from the SDG, $B(c_1 + c_2)$, which depends on the investment sum of the two players, is equal among them, irrespective of their investment levels in the SDG. (ii) The group reward from the fund, $R(e_1 + e_2)$, which depends on the donation sum of the two players, is shared between them, but each share may be unequal and based on merit. Fund-sharing may be modeled in various ways. In the model we examine a simple and intuitive case that the individual share of rewards is proportional to the degree of investment in the SDG, $c_i$.

We study the gradual evolution of populations on the two-dimensional trait space $(c, e)$, using the techniques of adaptive dynamics [8,9]. We first describe the fitness of a rare mutant with strategy $y = (c_y, e_y)$ among the residents with strategy $x = (c_x, e_x)$. This is called invasion fitness $F(x, y)$, defined by $F(x, y) = P(x, y) - P(x, x)$, where $P(x, y)$ denotes the expected payoff for the mutant with $y$. The expected payoff is given by

$$P(x,y) = B(c_x + c_y) + \frac{c_y}{c_x + c_y} R(e_x + e_y) - C(c_y) - S(e_y), \tag{1}$$

where the first term denotes the benefit from the SDG, and the second term, the individual reward which is proportional to the investment level, $c$.

For simplicity, we assume both the upper limits of $c_i$ and $e_i$ as 1. We assume quadratic payoff functions for the SDG and linear functions for the reward fund, as follows: $B(x) = \beta_2 x^2 + \beta_1 x$, $C(x) = \gamma_2 x^2 + \gamma_1 x$, $R(x) = rx$, and $S(x) = sx$. In the case of no rewarding, the monomorphic adaptive dynamics for quadratic cost and benefit functions have been fully analyzed. According to [2], the corresponding adaptive dynamics for monomorphic populations can have the unique equilibrium ("singular strategy") $Q = (c^*, 0)$ with $c^* = (\gamma_1 - \beta_1)/(4\beta_2 - 2\gamma_2)$. Singular strategy Q enters the $c$'s interval [0,1] and also is a (globally) convergence-stable, if and only if $\partial F/\partial c_y|_{y=x=(0,0)} = \beta_1 - \gamma_1 > 0$ and $\partial F/\partial c_y|_{y=x=(1,0)} = (4\beta_2 - 2\gamma_2) + (\beta_1 - \gamma_1) < 0$. Furthermore, Q is an evolutionary branching point at which a monomorphic population will diverge to two subpopulations across the point, if $\partial^2 F/\partial c_y^2|_{y=x=Q} = \beta_2 - \gamma_2 > 0$; or otherwise, an evolutionary merging point across which a dimorphic population can converge to a uniform population at the point [2,3].

## 3 Results and discussion

We shall demonstrate that evolutionary branching can profoundly affect the evolutionary fate of populations been stuck at a state in which all never reward: $e = 0$. It is easy to see that the homogeneous state of non-rewarding ($e = 0$) is globally convergence-stable, if $r - s < 0$; otherwise, that of full-rewarding ($e = 1$) is globally convergence-stable. Of special interest is the former case, in which individuals tend to decrease in the donation level in monomorphic populations. In the case singular strategy Q = ($c^*$, 0) can be globally convergence-stable for gradual co-evolution of investment and donation ($c$, $e$). According to numerical simulations (Fig. 1), indeed the monomorphic populations, starting with $e = 1$, evolves towards singular strategy Q (Fig. 1**a**). Voluntarily rewarding vanishes for the adaptive dynamics of monomorphic populations.

Subsequently, evolutionary branching on the investment trait, $c$, takes place. A cooperative branch ("C-branch") whose investment level is greater than $c^*$, on the one hand, moves towards $c = 1$. For the specific parameters, on the way C-branch also starts evolving with respect to the donation trait $e$ (Fig. 1**d**). At the same time, the averages of investment and payoff abruptly increase (Fig. 1**e**,**f**). On the other hand, defective branch ("D-branch") first evolves to smaller levels of $c$. When the evolution of C-branch is only on trait $e$, the evolution of D-branch turns into of the opposite direction, i.e., increase in $c$ (Fig. 1**c**). The population in the end consists of C-branch with full investment and donation and D-branch with intermediate investment level and no donation. In particular, voluntarily rewarding emerges for the adaptive dynamics of dimorphic populations.

The results show that evolutionary branching can significantly affect the average levels of payoff, investment in the SDG, and donation in the reward fund, by facilitating the emergence of pool reward. Evolutionary branching at Q consequently leads to discontinuous increases all in these three indexes. We note that without pool reward, gradual evolution of the investment level, $c$, can cause dimorphic diversification of the population into the extreme levels: $c = 0$ and $c = 1$. In the case it is known that evolutionary branching can only lead to negligible increase in the average cooperation and also to decrease in the average payoff [3].

So far, most of studies on the evolution of selective incentives have made much effort on punishment [10]. Discrete-strategy models have confirmed that costly punishment is not likely to emerge when rare, without any supportive mechanism, under the SDG-like conditions that cooperators and defectors can coexist [11,12]. How evolutionary branching affects the gradual evolution of punishment would be another fascinating question that deserves further works.

**Figure 1**

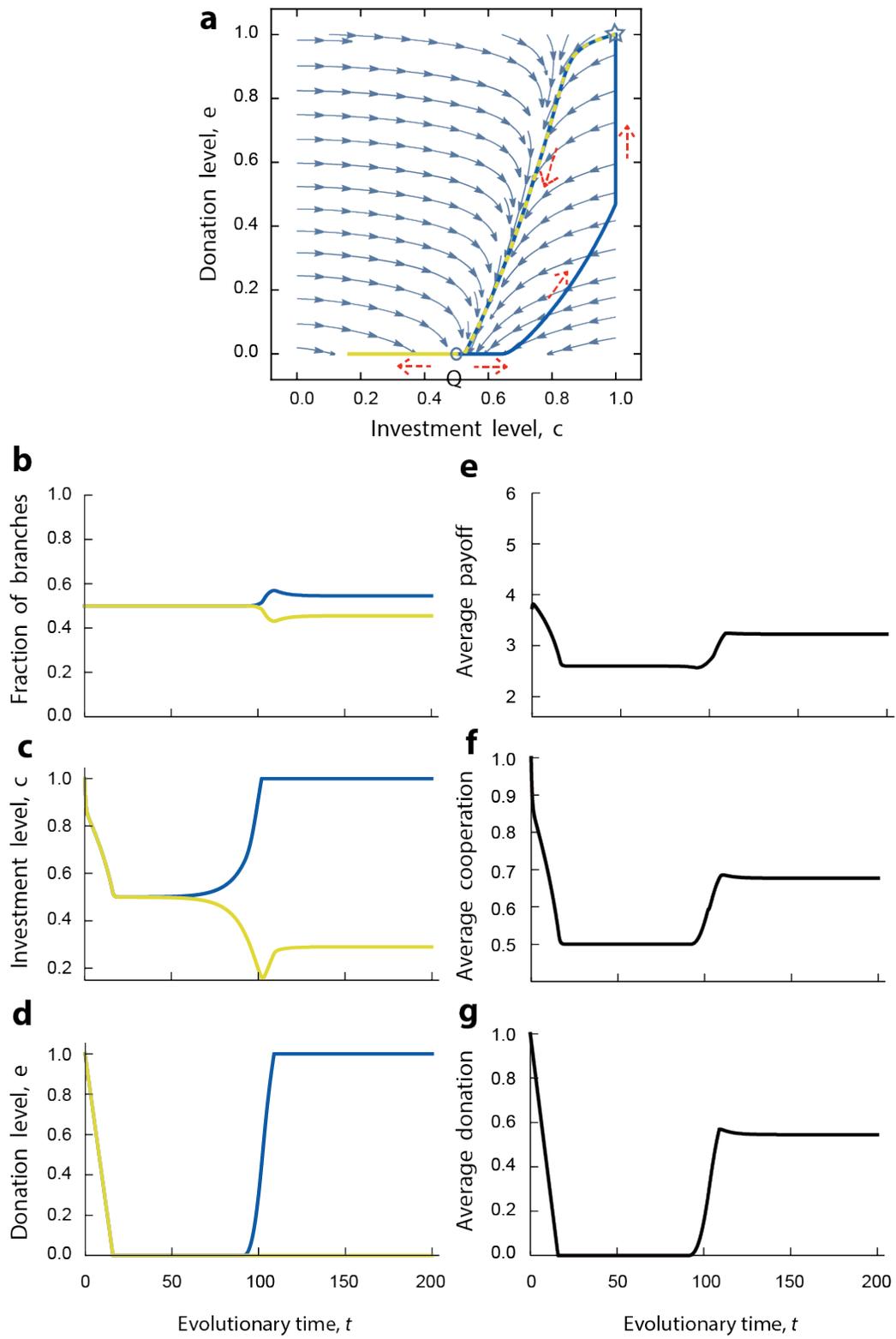

**Figure legend**

**Figure 1. Convergence, evolutionary branching, and emergence of voluntary reward.** The population starts its gradual evolution at the corner with $(c, e) = (1, 1)$ indicated by a star. According to the selection gradient (depicted by stream plot in **a**), the population first travels along a dashed curve and converges to boundary point Q (0.5, 0) indicated by an empty circle. Subsequently, evolutionary branching on trait $c$ occurs across Q around at $t = 50$ (**c**). A cooperative branch (dark blue; C-branch), on the one hand, moves to $c = 1$ and around at $t = 100$, also starts increasing in trait $e$ (**d**, **e**). At the same time, the average cooperation (**f**) and payoff (**e**) jump up. On the other hand, defective branch (light yellow; D-branch) first diverges to smaller levels of $c$ (**c**). When C-branch evolves along the line $c = 1$, the evolution of D-branch turns into of the opposite direction, i.e., increase in trait $c$. Eventually, the population becomes dimorphism consisting of C-branch with full investment and donation at (1, 1) and D-branch with intermediate investment and no donation at (0.29, 0).

To obtain the trajectory, I numerically solve the corresponding canonical equations for the adaptive dynamics [8] by using *Mathematica*. To allow transitions between monomorphism and dimorphism, I consider a population that initially consists of two patches whose locations are infinitesimally close to each other and fractions are equally 0.5. Other parameters: $B(x) = -1.4x^2 + 6x$, $C(x) = -1.6x^2 + 4.8x$, $R(x) = 1.75x$, and $S(x) = x$.